\documentclass[runningheads]{llncs}
\usepackage{butterma}
\idline{The final authenticated version is available online in the Proceedings of the 12th Symposium on Search-Based Software Engineering (SSBSE~2020) at Springer/SpringerLink.}

\usepackage{graphicx}
\usepackage{multirow}
\usepackage{url}

\usepackage{xspace} 
\newcommand{\tool}[1]{\textsc{#1}\xspace}
\newcommand{\evogfuzz}{\tool{EvoGFuzz}}
\newcommand{\activity}[1]{{Activity\,#1}\xspace}

\newcommand{\myparagraph}[1]{\noindent{\textbf{#1.}}}

\usepackage{mathtools}
\usepackage{amsmath}
\usepackage{nccmath}

\usepackage{subfigure}
\usepackage{dblfloatfix}

\usepackage{enumitem}
\usepackage{tcolorbox}

\usepackage{hyperref}
\usepackage{cite}

\begin{document}
\pagestyle{headings}
\setlength{\abovedisplayskip}{1pt}
\setlength{\belowdisplayskip}{2.1pt}

\title{Evolutionary Grammar-Based Fuzzing}

\author{Martin Eberlein
	\and
Yannic Noller
	\and
Thomas Vogel
	\and
Lars Grunske
}

\authorrunning{M. Eberlein et al.}

\institute{\textit{Software Engineering Group, Humboldt-Universit\"{a}t zu Berlin}, Berlin, Germany \\
\email{\{eberlema, yannic.noller, thomas.vogel, grunske\}@informatik.hu-berlin.de}	
}

\maketitle
\thispagestyle{electronic}
\begin{abstract}
A \emph{fuzzer} provides randomly generated inputs to a targeted software to expose erroneous behavior. To efficiently detect defects, generated inputs should conform to the structure of the input format and thus, grammars can be used to generate syntactically correct inputs. In this context, fuzzing can be guided by probabilities attached to competing rules in the grammar, leading to the idea of probabilistic grammar-based fuzzing. However, the optimal assignment of probabilities to individual grammar rules to effectively expose erroneous behavior for individual systems under test is an open research question. In this paper, we present \evogfuzz, an \textit{evo}lutionary \textit{g}rammar-based \textit{fuzz}ing approach to optimize the probabilities to generate test inputs that may be more likely to trigger exceptional behavior. The evaluation shows the effectiveness of \evogfuzz in detecting defects compared to probabilistic grammar-based fuzzing (baseline). Applied to ten real-world applications with common input formats (JSON, JavaScript, or CSS3), the evaluation shows that \evogfuzz achieved a significantly larger median line coverage for all subjects by up to 48\% compared to the baseline. Moreover, \evogfuzz managed to expose 11 unique defects, from which five have not been detected by the baseline.
\end{abstract}


\section{Introduction}

Software security vulnerabilities can be extremely costly \cite{richardson_csi_2008}.
Hunting down those issues has therefore been subject of intense research \cite{Godefroid2012Sage,Song2008Bitblaze,diffuzz}.
A typical example are internet browsers that combine a wide variety of interconnected components, using different interpreters and languages like JavaScript, Java, CSS, or JSON.
This makes web browsers extremely prone to exploiting the growing set of embedded parsers and interpreters to launch malicious attacks.
Hallaraker et al. \cite{hallaraker_detecting_2005} have shown that in particular the JavaScript interpreter, which is used to enhance the client-side display of web pages, is responsible for high-level security issues, allowing attackers to steal users' credentials and lure users into divulging sensitive information.
Unfortunately, due to the steady increase in complexity, interpreters become increasingly hard to test and verify.

Fuzzing \cite{Miller1990,Godefroid2020Fuzzing} has shown great success in finding vulnerabilities and erroneous behavior in a variety of different programs and software \cite{noauthor_afl_2018,noauthor_libfuzzer_2018}.
A \emph{fuzzer} generates random input data and enhances or mutates them to trigger potential defects or software vulnerabilities.
In general, fuzzing comes in various flavors: blackbox, whitebox, and greybox fuzzing \cite{Godefroid2020Fuzzing}.
While blackbox fuzzers have no knowledge about the internals of the application under test and apply random input generation, whitebox fuzzers can unleash the full power of program analysis techniques to use the retrieved context information to generate inputs.
Greybox fuzzing strikes a balance between these two cases: it employs a light-weight instrumentation of the program to collect some feedback about the generated inputs, which is used to guide the mutation process. This approach reduces the overhead significantly and makes greybox fuzzing an extremely successful vulnerability detection technique \cite{Bohme2017AFLGo}.
%
Nevertheless, greybox fuzzers still struggle to create semantically and syntactically correct inputs \cite{pham_smart_2019}.
The lack of the structural input awareness is considered to be the main limitation.
Since greybox fuzzers usually apply mutations on the bit level representation of an input, it is hard to keep a high level, syntactically correct structure.
Yet, to detect vulnerabilities deep inside programs, complex input files are needed.

Recently, Pavese et al. \cite{pavese_inputs_2018} presented an approach to generate test inputs using a grammar and a set of input seeds.
By using the input seeds to obtain a probabilistic grammar, Pavese et al. generate similar inputs to the seeds, or by inverting probabilities of the grammar, generate dissimilar inputs.
Similar input samples can be very useful, for instance, when learning from failure-inducing samples, while dissimilar inputs can be very useful for testing less common, and therefore less evaluated parts of a program.
We pick up this general idea of generating inputs based on a probabilistic grammar and propose \textit{evolutionary grammar-based fuzzing} (\evogfuzz), which combines the technique with an evolutionary optimization approach to detect defects and unwanted behavior in parsers and interpreters.
By using a probabilistic grammar, the fuzzer is able to generate syntactically correct inputs.
Furthermore, our concept of an evolutionary process is able to generate \textit{interesting} (i.e., failure-inducing inputs) and \textit{complex} input files (e.g., nested loops in JavaScript or nested data structures in JSON).
Utilizing the probabilistic grammar to generate new populations allows for good guiding properties.
By selecting the most promising inputs of a population and by learning and evolving the probabilistic grammar accordingly, essentially favoring specific production rules from the previous population, this process allows the directed creation of inputs towards specific features.
Additionally, \evogfuzz aims to be language and grammar independent to appeal to a broader testing community.

To examine the effectiveness of our approach, we implemented \evogfuzz as an extension of the tool by Pavese et al.~\cite{pavese_inputs_2018} and conducted experiments on several subjects for three common input languages and their parsers: JSON, JavaScript, and CSS3.
We compared \evogfuzz with the original approach and observed that within the same resource budget our technique can significantly increase the program coverage.
Moreover, \evogfuzz has been able to trigger more exception types (\evogfuzz 11 vs. the original approach 6). 

In summary, this paper makes the following contributions:
\vspace{-1em}
\begin{itemize}
\item We propose an evolutionary grammar-based fuzzing approach (\evogfuzz) that combines the concept of probabilistic grammars and evolutionary algorithms to generate test inputs that trigger defects and unwanted behavior.
\item We implement \evogfuzz as an extension of an probabilistic grammar-based fuzzer~\cite{pavese_inputs_2018} and the ANTLR parser generator.
\item We demonstrate the effectiveness of \evogfuzz on ten real-world examples across multiple programming languages and input formats, and provide a comparison with the original approach~\cite{pavese_inputs_2018}.
\end{itemize}

\section{Related Work}
\evogfuzz focuses on the generation of test inputs to reveal defects and unwanted behavior.
Existing approaches in this area can be separated in \textit{search-based}, \textit{generative}, \textit{learning-based}, and \textit{combined} techniques~\cite{Anand2013SurveyTesting,Orso2014SoftwareTesting}.

\myparagraph{Search-based input generation}
Test input generation can be formulated as a \textit{search} problem to be solved by meta-heuristic search~\cite{Harman2012SBSE,fuzzingbook2019}.
A simple way is to \textit{randomly} generate inputs, as employed in the original work on fuzzing by Miller et al.~\cite{Miller1990}.
More sophisticated random testing strategies are directed by \textit{feedback}~\cite{Pacheco2007Randoop}.
Evolutionary search applies fitness functions to select promising inputs, while the inputs are generated by mutating an initial population.
Recent advances in \textit{fuzzing} show the strength of such search-based techniques~\cite{noauthor_afl_2018,Bohme2016AFLFast,Lemieux2018FairFuzz}.
One of the most popular greybox fuzzers is \tool{AFL}~\cite{noauthor_afl_2018} that applies a genetic algorithm guided by coverage information.
While these techniques can successfully generate error-revealing inputs, they miss required information about a program to generate syntactically and semantically correct inputs~\cite{pham_smart_2019,Wang2019Superion}.

\myparagraph{Generative input generation}
Hanford~\cite{Hanford1970} introduced grammar-based test generation with his \textit{syntax machine}.
Recent advances in \textit{grammar-based fuzzing} pick up this idea and use a grammar to generate inputs that are syntactically correct~\cite{Godefroid2008GrammarWhiteBoxFuzzing,Holler2012FuzzingCodeFragments}.
The main focus of grammar-based fuzzers are parsers and compilers~\cite{Yang2011CSmith,Holler2012FuzzingCodeFragments}.
Having grammar production rules augmented with probabilities (aka \textit{probabilistic grammars}) allows to generate inputs based on rule prioritization.
Pavese et al.~\cite{pavese_inputs_2018} employ this notion: they take an input grammar, augment it with probabilities and generate structured inputs that represent common or very uncommon inputs.
In general, generative approaches require the input grammar or language specification, which might not always be a available or accurate enough.
Therefore, H{\"o}schele and Zeller~\cite{Hoschele2017AUTOGRAM2} proposed input grammar mining.

\myparagraph{Learning-based input generation}
In addition to grammar mining, machine learning is increasingly applied for software testing~\cite{Godefroid2017LearnFuzz,Cummins2018,Liu2017}. 
Those techniques learn input structures from seed inputs and use them to generate new testing sequences.
They target web browsers~\cite{Godefroid2017LearnFuzz}, compilers~\cite{Cummins2018}, and mobile apps~\cite{Liu2017}. 

\myparagraph{Combined techniques}
Recently, a lot of research efforts focus on the combination of grammar-based and coverage-based fuzzing (CGF) with the goal to use the grammar to generate valid inputs but to use CGF to further explore the input space.
Le et al.~\cite{Le2019Saffron} propose a fuzzer that generates inputs for the worst-case analysis.
While they leverage a grammar to generate seed inputs for a CGF, they continuously complement/refine the grammar.
Atlidakis et al.~\cite{Atlidakis2020} propose \tool{Pythia} to test REST APIs.
They learn a statistical model that includes the structure of specific API requests as well as their dependencies across each other.
While \tool{Pythia}'s mutation strategies use this statistical model to generate valid inputs, coverage-guided feedback is used to select inputs that cover new behavior.
Other fuzzing works aim to incorporate grammar knowledge within their mutation strategies~\cite{pham_smart_2019,Wang2019Superion}. 
Similarly to \tool{Pythia}, we use seed inputs to generate an initial probabilistic grammar.
However, with every iteration we retrieve new probabilities for the grammar while also mutating these probabilities, which enables evolution of the grammar and a broad exploration of the input~space.

\section{Evolutionary Grammar-Based Fuzzing (\evogfuzz)}\label{sec:approach}

In this section, we will present \evogfuzz, a language-independent evolutionary grammar-based fuzzing approach to detect defects and unwanted behavior. 

\begin{figure}[t]
\centering
  \includegraphics[width=1\textwidth]{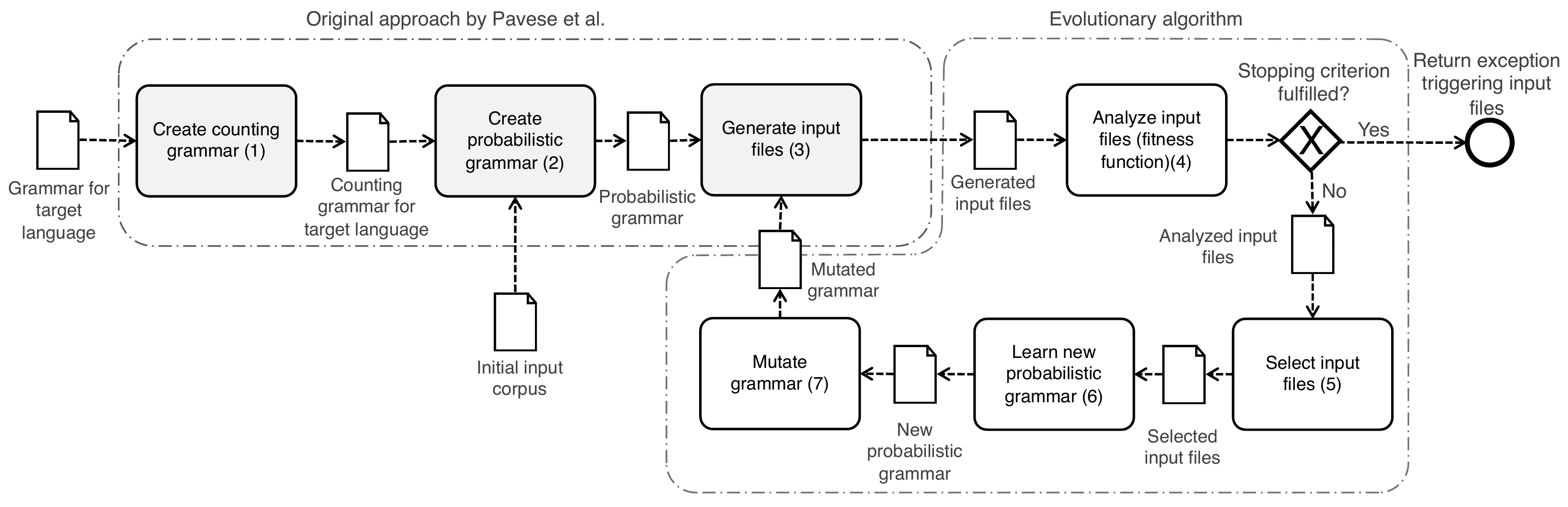}
  \caption{Overview of \evogfuzz.}
  \label{fig:overview}
\end{figure}

The key idea of \evogfuzz is to combine probabilistic grammar-based fuzzing and evolutionary computing. This combination aims for directing the probabilistic generation of test inputs toward ``interesting'' and ``complex'' inputs. The motivation is that ``interesting'' and ``complex'' inputs more likely reveal defects in a software under test~(SUT).
For this purpose, we extend an existing probabilistic grammar-based fuzzer~\cite{pavese_inputs_2018} with an evolutionary algorithm (Figure~\ref{fig:overview}).

Similarly to the original fuzzer, \evogfuzz requires a correctly specified \textit{grammar for the target language}, that is, the input language of the SUT.
%
From this grammar, we create a so-called \textit{counting grammar} (\activity{1} in Figure~\ref{fig:overview}) that describes the same language but allows us to measure how frequently individual choices of production rules are used when parsing input files of the given language.
%
Thus, the counting grammar allows us to learn a \textit{probabilistic grammar} from a sampled \textit{initial corpus of inputs} (\activity{2}). 
Particularly, we learn the probabilities for selecting choices of production rules in the grammar, which correspond to the relative numbers of using these choices when parsing the initial corpus.
%
Consequently, we use the probabilistic grammar to generate more input files that resemble features of the initial corpus, that is, ``more of the same''~\cite{pavese_inputs_2018} is produced (\activity{3}).
Whereas this activity is the last step of the approach by Pavese et al.~\cite{pavese_inputs_2018}, it is the starting point of the evolutionary process in \evogfuzz as it generates an initial population of test inputs. An individual of the population therefore corresponds to a single input file for the SUT.

%
The evolutionary algorithm of \evogfuzz starts a new iteration with analyzing each individual using a fitness function that combines feedback and structural scores (\activity{4}). By executing the SUT with an individual as input, the feedback score determines whether the individual triggers an exception. If so, this input is considered as an ``interesting'' input. The structural score quantifies the ``complexity'' of the individual.
If the stopping criterion is fulfilled (e.g., a time budget has been completely used), the exception-triggering input files are returned.
%
Otherwise, the ``interesting'' and most ``complex'' individuals are selected for evolution (\activity{5}).
%
The selected individuals are used to learn a \textit{new probabilistic grammar}, particularly the probability distribution for the production rules similarly to Activity~2 that, however, used a sampled initial corpus of inputs (\activity{6}). Thus, the new probabilistic grammar supports generating ``more of the same'' interesting and complex inputs.
%
To mitigate a genetic drift toward specific features of the selected individuals, we mutate the new probabilistic grammar by altering the probabilities for randomly chosen production rules (\activity{7}).
Finally, using the mutated probabilistic grammar, we again generate new input files (\activity{8}) starting the next evolutionary iteration.

Assuming that inputs similar to ``interesting'' and ``complex'' inputs more likely reveal defects in the SUT, \evogfuzz guides the generation of inputs toward ``interesting'' and ``complex'' inputs by iteratively generating, evaluating, and selecting such inputs, and learning (updating) and mutating the probabilistic grammar. 
In contrast to typical evolutionary algorithms, \evogfuzz does not directly evolve the individuals (test input files) by crossover or mutation but rather the probabilistic grammar whose probabilities are iteratively learned and mutated.
In the following, we will discuss each activity of \evogfuzz in detail.

\subsection{Probabilistic Grammar-Based Fuzzing (Activities~1--3)}\label{sec:pavese}

Pavese et al.~\cite{pavese_inputs_2018} have proposed probabilistic grammar-based fuzzing to generate test inputs that are similar to a set of given inputs. Using a (context-free) grammar results in syntactically correct test inputs being generated. However, the production rules of the grammar are typically chosen randomly to generate inputs, which does not support influencing the features of the generated inputs. To mitigate this situation, Pavese et al. use a probabilistic grammar, in which probabilities are assigned to choices of production rules. The distribution of these probabilities are learned from a sample of inputs. Consequently, test inputs produced by a learned probabilistic grammar are therefore similar to the sampled inputs. Pavese et al. call this idea ``more of the same''~\cite{pavese_inputs_2018} because the produced inputs share the same features as the sampled inputs.

Technically, a given context-free grammar for the input language of the SUT is transformed to a so-called counting grammar by adding a variable to each choice of all production rules (\activity{1}). Such a variable counts how often its associated choice of a production rule is executed when parsing a given sample of inputs. Knowing how often a production rule is executed in total during parsing, the probability distribution of all choices of the rule is determined. 
Thus, using the counting grammar to parse a sample of $n_{sample}$ input files, the variables of the grammar are filled with values according to the executed production rules and their choices (\activity{2}). This results in a probabilistic grammar, in which a probability is assigned to each choice of a production rule.
Using this probabilistic grammar, we can generate new input files that resemble features of the sampled input files since both sets of input files have the same probability distribution for the production rules (\activity{3}).
Thus, \evogfuzz uses the approach by Pavese et al. to initially learn a probabilistic grammar from $n_{sample}$ input files (Activities\,1 and~2), and to generate $|P|$ new input files for the (initial) population $P$ (\activity{3}), which starts an evolutionary iteration.

\subsection{Evolutionary Algorithm (Activities~4--8)}\label{sec:evo}

The evolutionary algorithm of \evogfuzz evolves a population of test input files by iteratively 
(i)~analyzing the fitness of each individual, 
(ii)~selecting the fittest individuals, 
(iii)~learning a new probabilistic grammar based on the selected individuals, 
(iv)~mutating the learned grammar, and 
(v)~generating new individuals by the mutated grammar that form the population for the next iteration.

\myparagraph{Analyze individuals (\activity{4})}
Our goal is to evolve individuals toward ``complex'' and ``interesting'' test inputs as such inputs more likely detect defects and unwanted behavior.
To achieve this goal, we use a fitness function that quantifies both aspects, the complexity and whether an input is of interest.

Concerning complexity, we focus on the structure of test input files assuming complex structures (e.g., nested loops in JavaScript) have a higher tendency to reveal uncommon behavior in the SUT (e.g., a JavaScript parser) than simple ones. 
However, we can make only few assumptions about the complexity of input files since \evogfuzz is language independent and thus, it has no semantic knowledge about the language of test inputs besides the grammar.
Thus, we can only rely on \textit{generic} features of test input files and grammars to quantify the complexity of an input file.
A straightforward and efficient metric to use would be the \textit{length} of an input file in terms of the number of characters contained by the file.
However, a fitness function maximizing file length would favor production rules that produce terminals being longer strings (e.g., ``true'' or ``false'' in JavaScript) over rules that produce more expansions through non-terminals to obtain complex structures (e.g., ``if'' branches or ``for'' loops).
To mitigate this effect, we further use \textit{the number of used expansions to derive an input file} because using more expansions to generate an input file makes the input file more complex. 
Thus, we build the \textit{ratio} of the number of \textit{expansions} to the \textit{length} of an input file $x$ to favor input files that were produced by more expansions and to punish lengthy input files that contain long strings of characters. 
Depending on the language of the input files, this ratio can be controlled by the parameter~$\lambda$. 
\begin{ceqn}
	\begin{align}
	ratio(x) = \frac{expansions(x)}{\lambda \times length(x) }
	\label{eq:ratio}
	\end{align}
\end{ceqn}
Using this ratio, we score the structure of an input file $x$ by multiplying the ratio and the number of expansions to put more weight on the expansions while a good ratio ($>$1) increases the score.
\begin{ceqn}
	\begin{align}
	score_{structure}(x)= ratio(x) \times expansions(x)
	\label{eq:improvedstructure}
	\end{align}
\end{ceqn}
Benefits of this score are its efficient computation and independence of the input language, although controlling $\lambda$ allows accommodation of a specific language.

Concerning the ``interesting'' inputs, we rely on the feedback from executing the SUT with a concrete input $x$. Being interested in revealing defects in the SUT, we observe whether $x$ triggers any exception during execution. If so, such an input will be assigned the maximum fitness and favored over all other non-exception triggering inputs.
This results in a feedback score for an input file~$x$:
\begin{ceqn}
	\begin{align}
	score_{feedback}(x)=\begin{cases}
	\infty & \text{if $x$ triggers any exception} \\ 
	0 & \text{otherwise} 
	\end{cases}
	\label{eq:feedback}
	\end{align}
\end{ceqn}
Moreover, \evogfuzz keeps track of all exception-triggering inputs throughout all iterations as it returns these inputs at the end of the evolutionary search.

Finally, we follow the idea by Veggalam et al. \cite{veggalam_ifuzzer_2016} and combine the structural and feedback scores to a single-objective fitness function to be maximized:
\begin{ceqn}
	\begin{align}
	fitness(x)= score_{feedback}(x) + score_{structure}(x)
	\label{eq:fitness}
	\end{align}
\end{ceqn}
Using this fitness function, all $|P|$ input files generated by the previous activity (\activity{3}) are analyzed by executing them and computing their fitness.

\myparagraph{Select individuals (\activity{5})}
Based on the fitness of the $|P|$ input files, a strategy is needed to select the most promising files among them that will be used for further evolution. 
To balance selection pressure, \evogfuzz uses elitism~\cite{du_elitism_2018} and tournament selection~\cite{miller_genetic_1995}.
By elitism, the top $e_{rate}\%$ of the $|P|$ input files ranked by fitness are selected.
Additionally, the winners of $n_{tour}$ tournaments of size $s_{tour}$ are selected. The $s_{tour}$ participants of each tournament are randomly chosen from the remaining $(100-e_{rate})\%$ of the $|P|$ input files.
In contrast to typical evolutionary algorithms, the selected individuals are not directly evolved by crossover or mutation, but they are used to learn a new probabilistic grammar.

\myparagraph{Learn new probabilistic grammar (\activity{6})}
The selected input files are the most promising files of the population and they help in directing the further search toward ``complex'' and ``interesting'' inputs. Thus, these files are used to learn a new probabilistic grammar, particularly the probability distributions for all choices of production rules, by parsing them (cf.~\activity{2} that learns a probabilistic grammar, however, from a given sample of input files). Thus, the learned probability distributions reflect features of the selected input files, and the corresponding probabilistic grammar can produce more input files that resemble these features. 
But beforehand, \evogfuzz mutates the learned~grammar.

\myparagraph{Mutate grammar (\activity{7})}
We mutate the learned probabilistic grammar to avoid a genetic drift~\cite{wright_evolution_1929} toward specific features of the selected individuals. With such a drift, the grammar would generate only input files with specific features exhibited by the selected individuals from which the grammar has been learned. Thus, it would neglect other potentially promising, yet unexplored features.
Moreover, mutating the grammar maintains the diversity of input files being generated, which further could prevent the search from being stuck in local optima. 
In contrast to typical evolutionary algorithms, we do not mutate the individuals directly for two reasons. First, mutating an input file may result in a syntactically invalid file (i.e., the file does not conform to the given grammar).
Second, a stochastic nature of the search is already achieved by using a \textit{probabilistic} grammar to generate input~files.

Therefore, we mutate the learned probabilistic grammar by altering the probabilities of individual production rules. The resulting mutated grammar produces syntactically valid input files whose features are similar to the selected individuals but that may also exhibit other unexplored features. 
For instance, a mutation could enable choices of production rules in the grammar that have not been used yet to generate input files because of being tagged so far with a probability of 0 that is now mutated to a value larger than~0. This increases the genetic variation.

For a single mutation of a probabilistic grammar, we choose a random production rule with $n$ choices for expansions from the grammar. For each choice, we recalculate the probabilities $p_{i}$ by selecting a random probability~$r_{i}$ from $(0,1]$---we exclude $0$ to enable all choices by assigning a probability larger than zero---and normalizing $r_{i}$ with the sum of all of the $n$ probabilities to ensure $\sum_{i=1}^{n} p_{i} = 1$ (i.e., the individual probabilities of all choices of a production rule sum up to~1).
Thus, a probability $p_{i}$ for one choice is calculated as follows:
\begin{ceqn}
\begin{align}
	p_{i}=  \frac{r_{i}}{\sum_{j=1}^{n} r_{j} }
\end{align}
\end{ceqn}

Finally, \evogfuzz allows multiple of such mutations ($n_{mut}$ many) of a probabilistic grammar in one iteration of search by performing each mutation independently from the other~ones.

\myparagraph{Generate input files (\activity{3})}
Using the learned and mutated grammar, \evogfuzz generates $|P|$ new input files that resemble features of the recently selected input files but still diverge due to the grammar mutation. With the newly generated input files, the next iteration of the evolutionary process begins.

\section{Evaluation}

In this section, we evaluate the effectiveness of \evogfuzz by performing experimentation on ten real-world applications.\footnote{Data and code artifacts are available here: \url{https://doi.org/10.5281/zenodo.3961374}}
We compare \evogfuzz to a baseline being the original approach by Pavese et al.~\cite{pavese_inputs_2018} (i.e., probabilistic grammar-based fuzzing), and ask the following research questions:
\begin{center}
\vspace{-1em}
\begin{minipage}[t]{0.92\textwidth}
\begin{enumerate}[start=1,label={\bfseries RQ\arabic*}]
\item Can evolutionary grammar-based fuzzing achieve a higher code coverage than the baseline?
\item Can evolutionary grammar-based fuzzing trigger more exception types than the baseline?
\end{enumerate}
\end{minipage}
\vspace{-1em}
\end{center}

\subsection{Evaluation Setup}
To answer the above research questions, we conducted an empirical study, in which we analyze the achieved line coverage and the triggered exception types.
Line or code coverage \cite{miller_systematic_1963} is a metric counting the unique lines of code of the targeted parser (i.e., the SUT) that have been executed during a test.

In order to examine the effectiveness of \evogfuzz we evaluate our approach on the same test subjects that Pavese et al. have originally covered with their proposed probabilistic grammar-based fuzzing approach. These test subjects require three, in complexity varying input formats, namely \emph{JSON}, \emph{JavaScript}, and \emph{CSS3}. ARGO, Genson, Gson, JSONJava, JsonToJava, MinimalJson, Pojo, and json-simple serve as the JSON parsers, whereas Rhino and cssValidator serve as the JavaScript and CSS parser, respectively. All parsers are widely used in browsers and web applications. A further description of all subjects along with their grammars can be found in the work of Pavese et al \cite{pavese_inputs_2018}. All experiments have been performed on a virtual machine with Ubuntu 20.04 LTS featuring a Quad-Core 3GHz Intel(R) CPU with 16 GB RAM.

\subsection{Research Protocol}

Giving both approaches the same starting conditions, we considered the same randomly selected input files from Pavese et al. to create the initial probabilistic grammar.
The baseline uses this probabilistic grammar to generate ``more of the same'' inputs, whereas \evogfuzz uses this grammar to generate the initial population followed by executing its evolutionary algorithm.
In our evaluation, we observe the performance of both approaches for all subjects over a time frame of 10 minutes, that is, each approach runs for 10 minutes to test one subject.

For \evogfuzz, a population consists of 100 individuals ($|P|=100$) and one mutation of the grammar ($n_{mut}=1$) is performed in each iteration of the search. The elitism rate $e_{rate}$ is set to~$5\%$, and for each generation ten tournaments of size ten were held ($n_{tour}=10$ and $s_{tour}=10$). 
In the fitness function, $\lambda$ is set to $1.5$ for JSON and $2.0$ for JavaScript and CSS.
Since the goal is to find exceptions, we configured the baseline to perform iterations of generating and executing 100 ``more of the same'' input files for 10 minutes. 
After 10 minutes the baseline and \evogfuzz return all found exceptions and the exception-triggering test inputs.
For each test subject and approach, we repeated these experiments 30 times.

\subsection{Experimental Results}
Figures \ref{fig:argo} to \ref{fig:cssValidator} show the coverage results for the ten subjects. 
For each subject, we plot a chart showing the comparison of \evogfuzz and the baseline with regard to the achieved line coverage.
The vertical axis represents the achieved line coverage in percent, and the horizontal axis represents the time in seconds (up to 600 seconds = 10 minutes).
The \textit{median} runs for both approaches are highlighted, with all individual runs being displayed in the background. 

\myparagraph{RQ1 - Line coverage}
To answer RQ1, we compare the line coverage achieved by both approaches.
In particular, we investigate whether \evogfuzz achieves at least the same percentage of line coverage than the baseline.
Figures~\ref{fig:argo} to~\ref{fig:jsonsimple} show the results for the JSON parsers, and Figures~\ref{fig:rhino} and~\ref{fig:cssValidator} show the results for the JavaScript and CSS3 parser, respectively.

\newcommand{\subfigwidth}{48mm}

\newcommand{\mysubfig}[2]{%
\vspace{-2mm}
\subfigure{%
		\includegraphics[width=\subfigwidth]{graphics/#2}
		#1 
}%
}

\begin{figure}[!h]
	\centering
	\vspace{-2mm}
	\mysubfig{\label{fig:argo}}{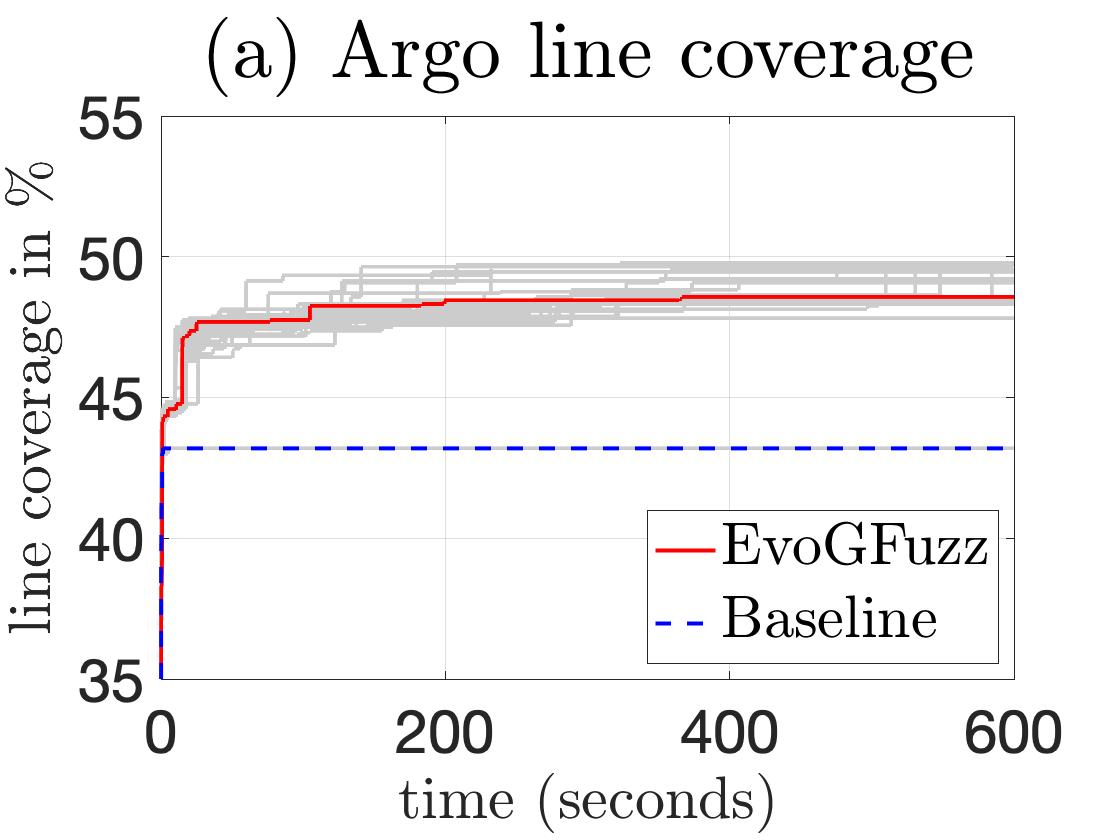}
	\mysubfig{\label{fig:genson}}{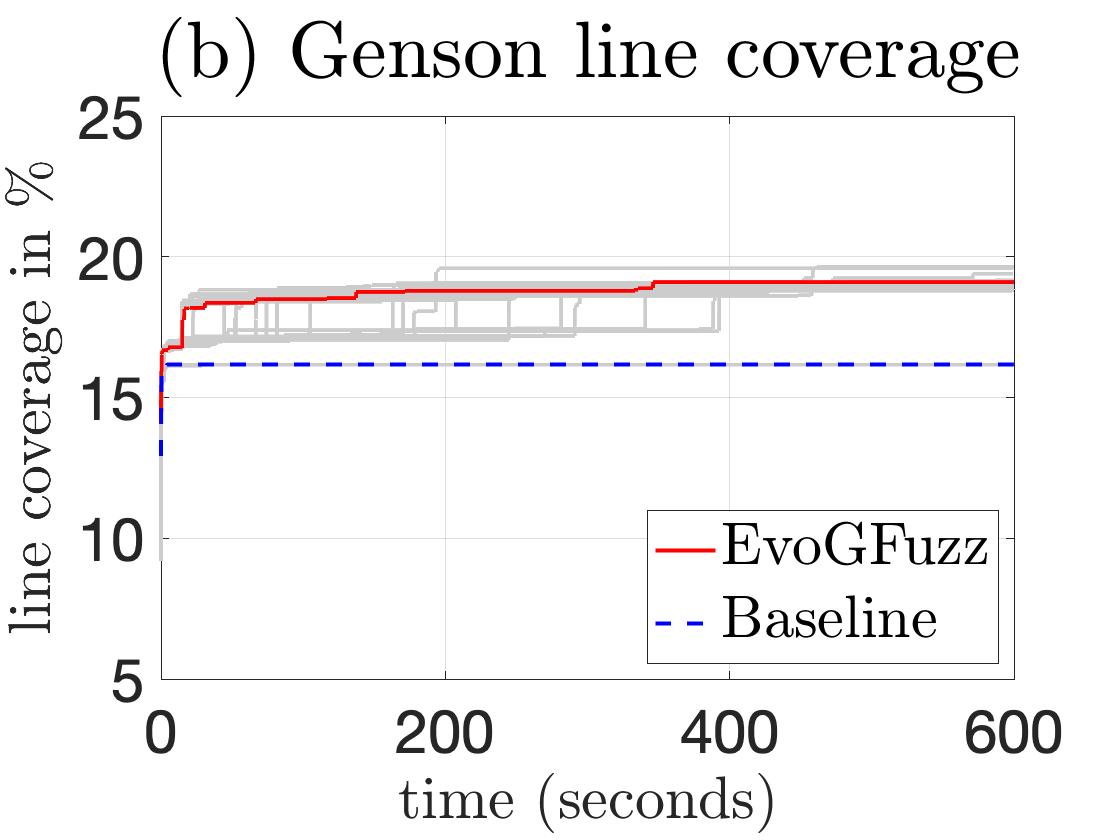}
	\mysubfig{\label{fig:gson}}{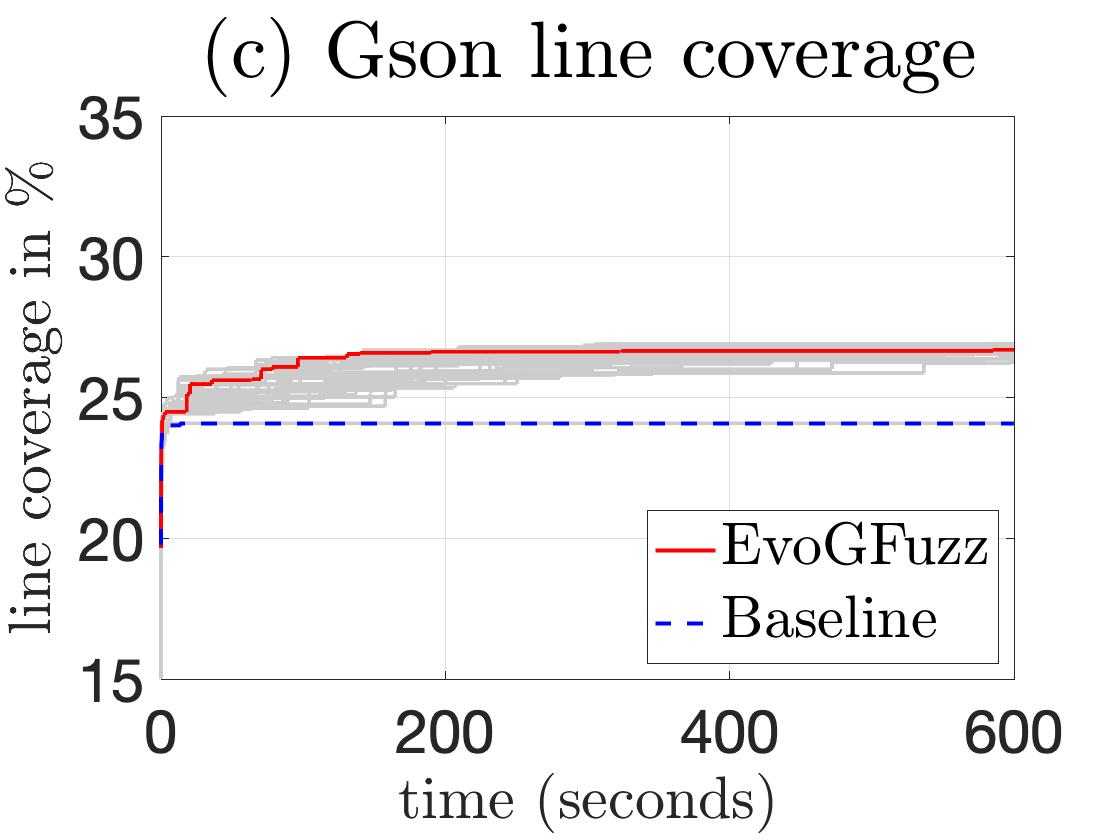}
	\mysubfig{\label{fig:jsonJava}}{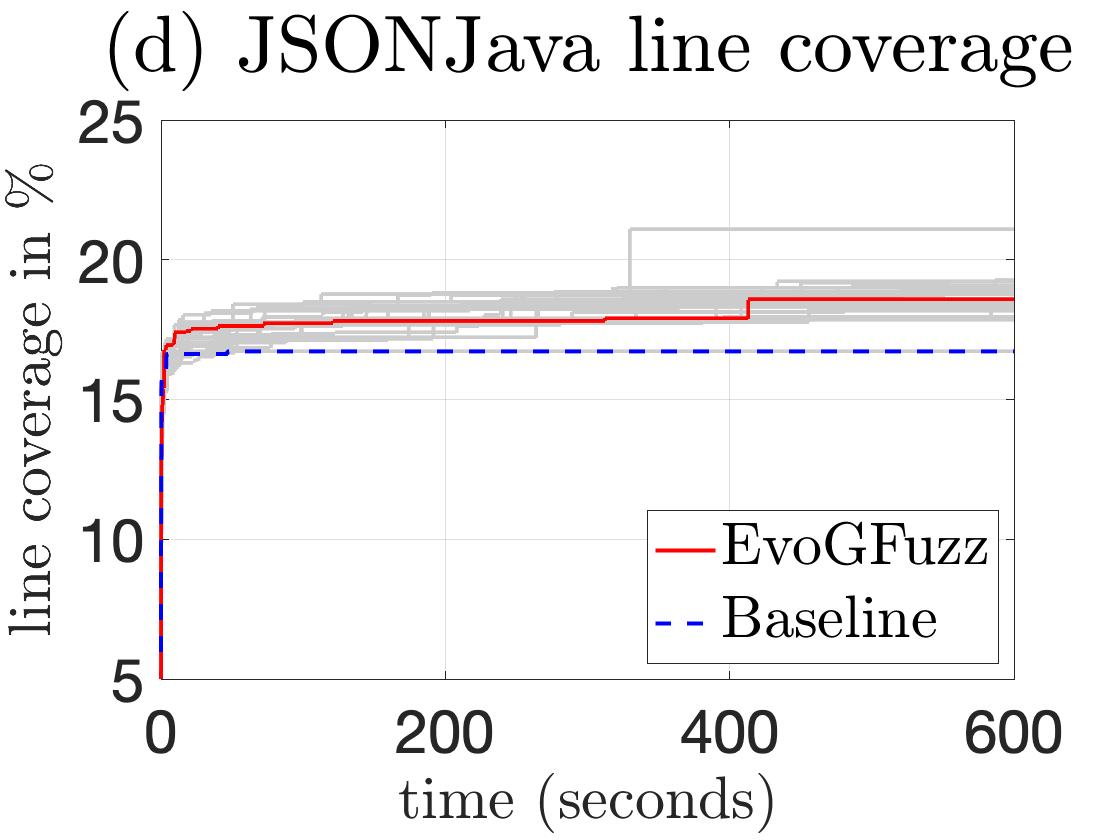}
	\mysubfig{\label{fig:jsonToJava}}{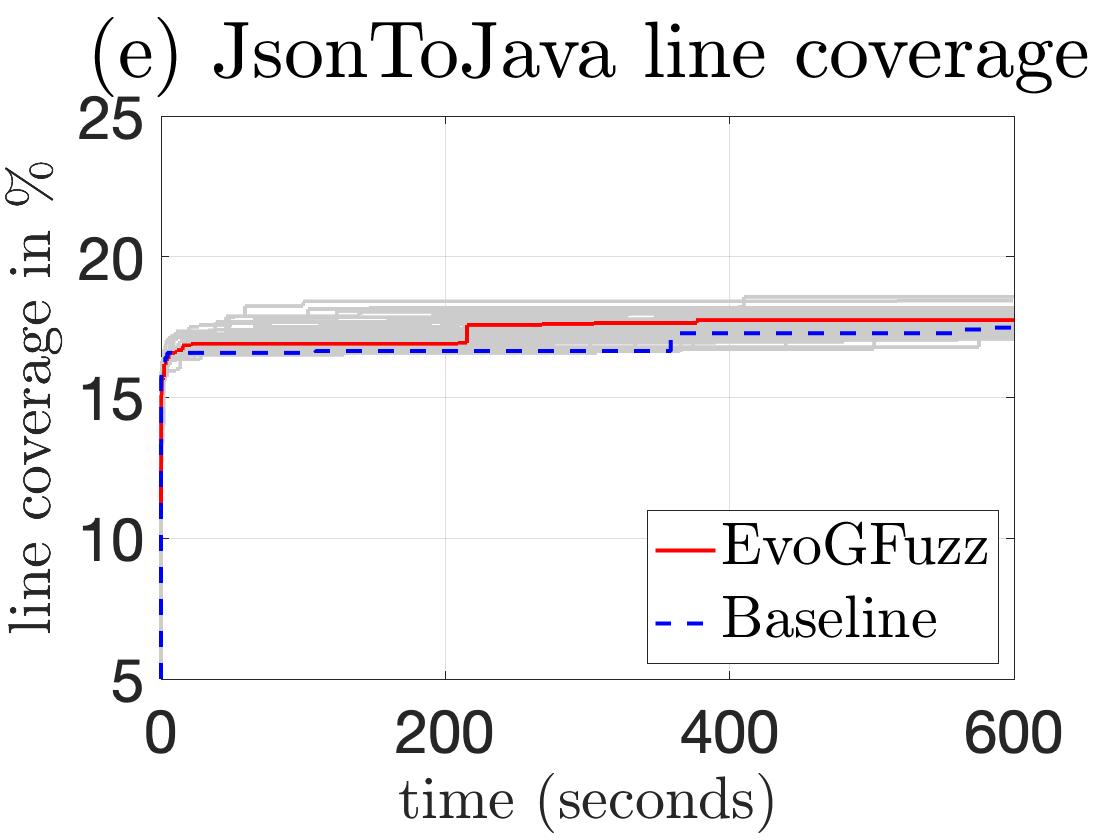}
	\mysubfig{\label{fig:minimalJson}}{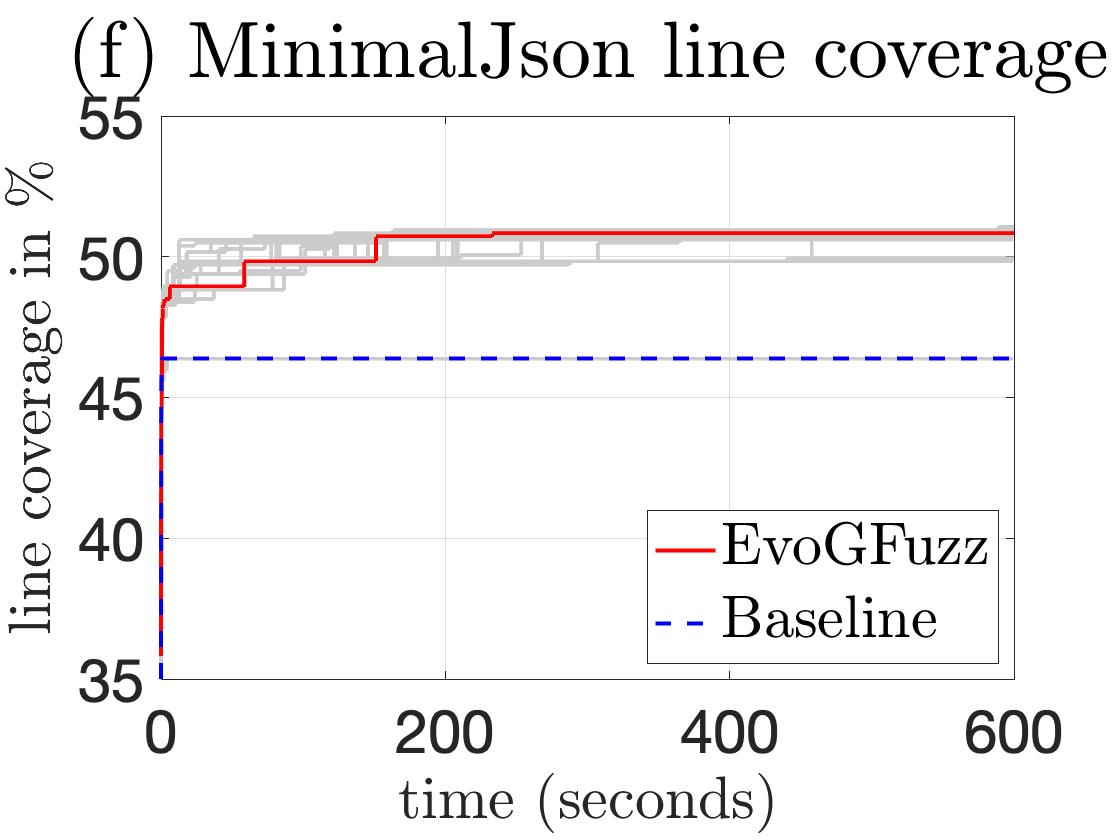}
	\mysubfig{\label{fig:pojo}}{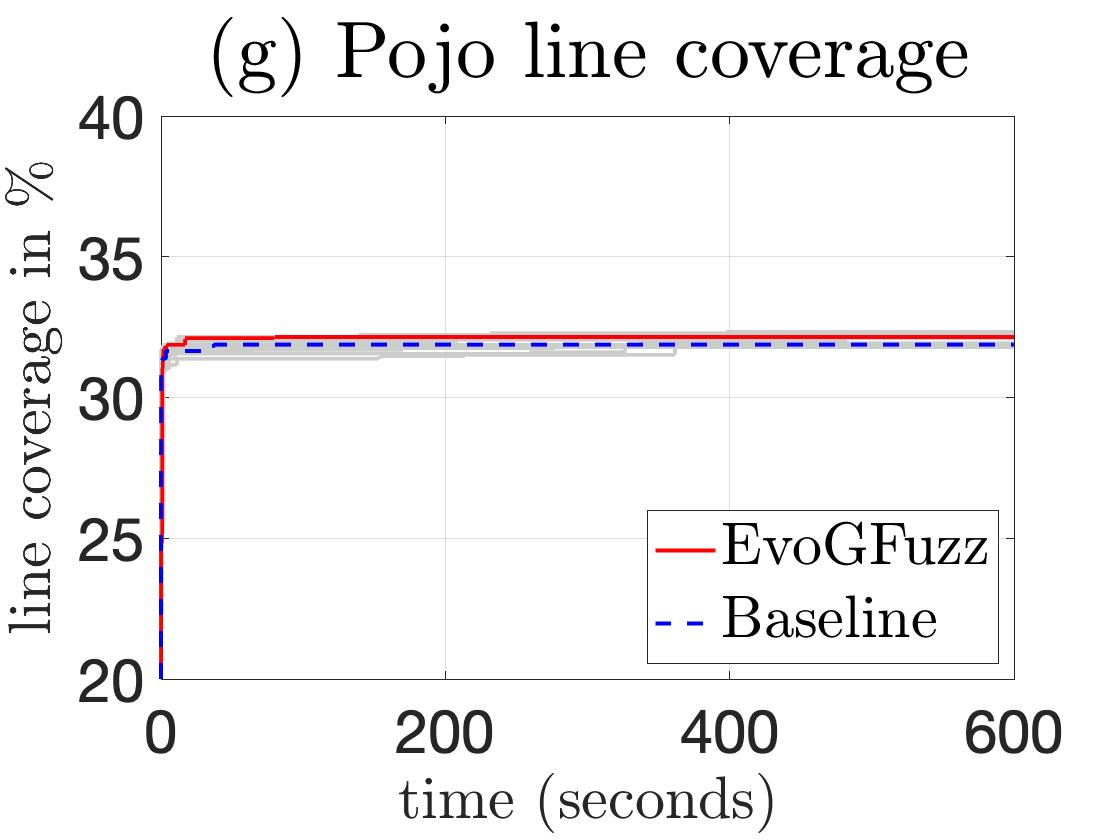}	
	\mysubfig{\label{fig:jsonsimple}}{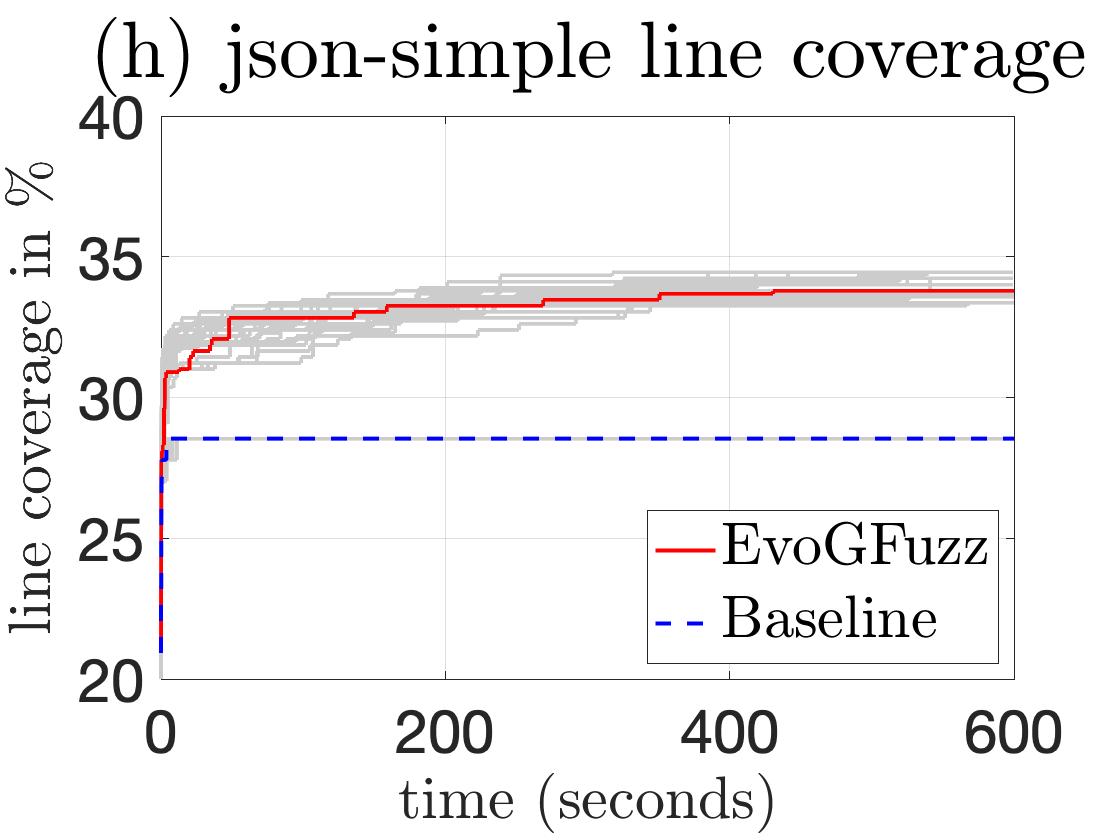}	
	\mysubfig{\label{fig:rhino}}{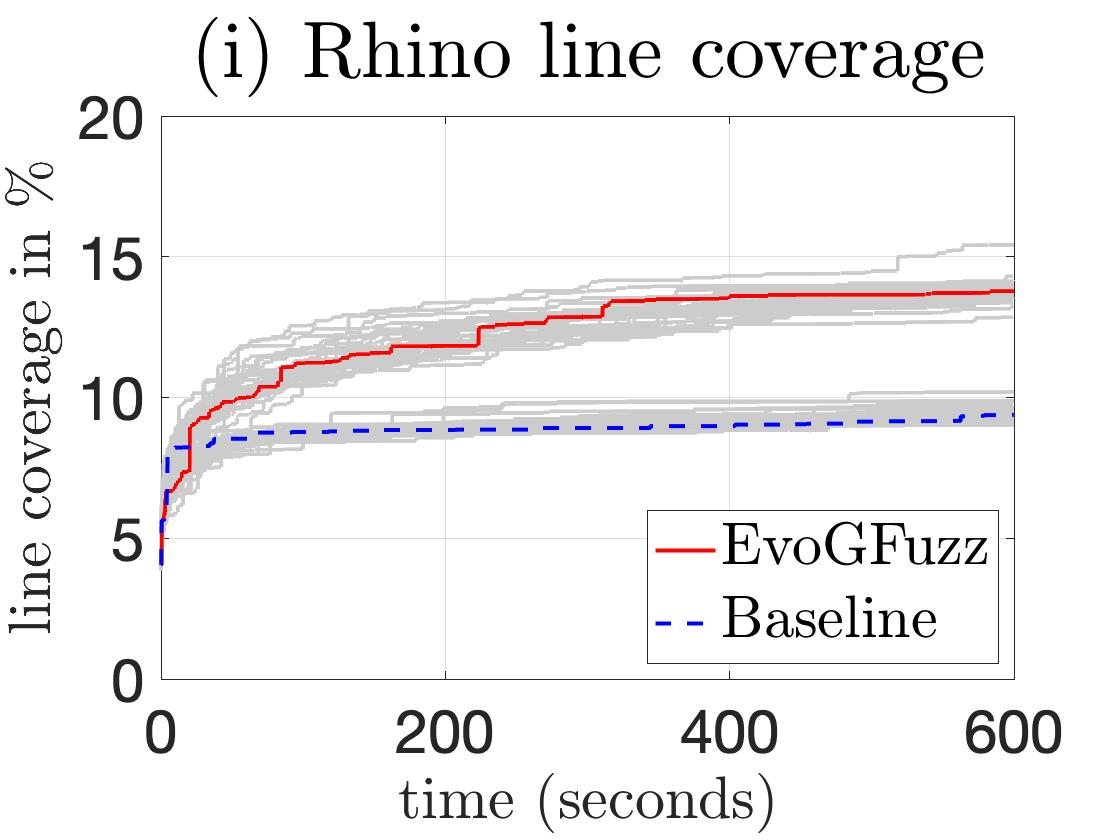}	
	\mysubfig{\label{fig:cssValidator}}{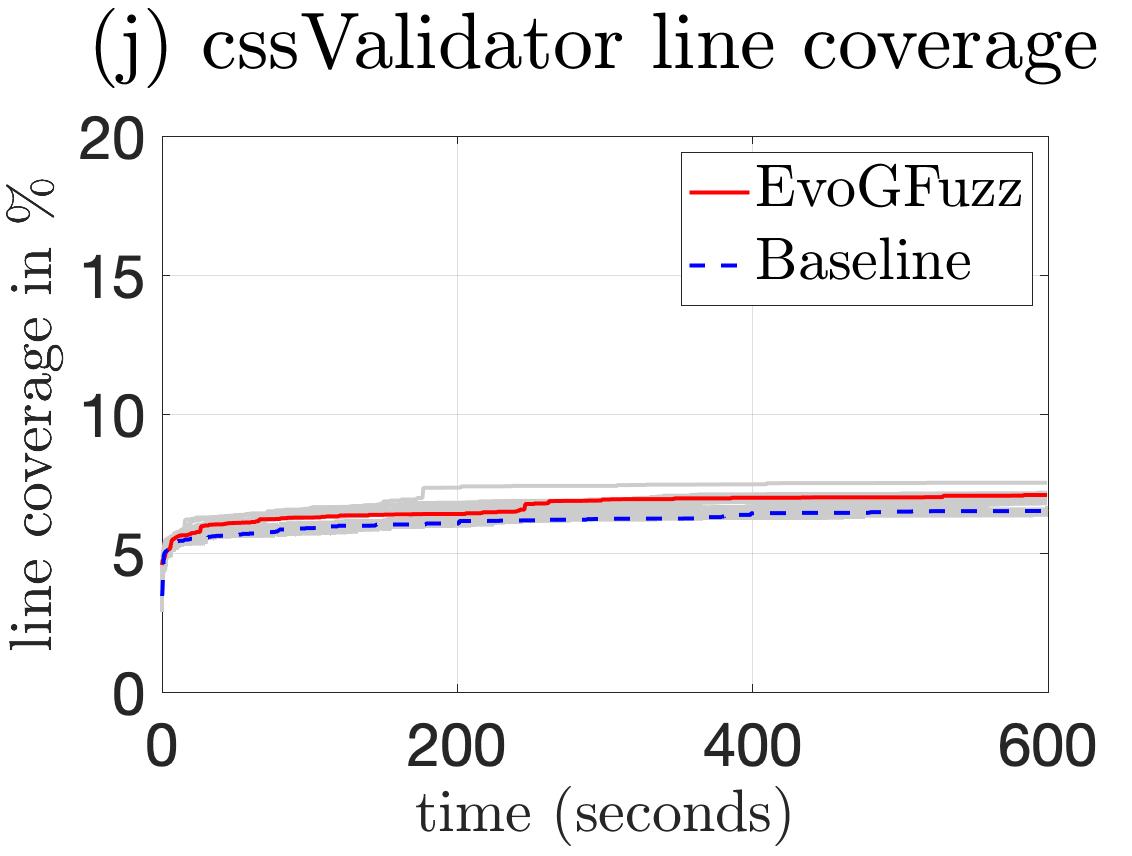}
	\caption{Median and Raw Line Coverage Results for the Ten Subjects.}
	\label{fig:linecoverage}
\end{figure} 

The results show that \evogfuzz improves the coverage for all subjects and is able to increase the median line coverage 
for JSON by up to 18.43\% (json-simple, Fig. \ref{fig:jsonsimple}), 
for JavaScript by up to 47.93\% (Rhino, Fig. \ref{fig:rhino}), and 
for CSS3 by up to 8.45\% (cssValidator, Fig. \ref{fig:cssValidator}). 
These numbers are also listed in the column ``Median increase'' of Table~\ref{table:evaluation}.

The detailed investigation of Figures~\ref{fig:argo} to~\ref{fig:jsonsimple} shows that for almost all JSON parsers both approaches eventually reach a plateau with regard to the achieved line coverage.
The baseline reaches this plateau relatively early in the input generation process: there is no further improvement after only approximately 10 seconds.
For \evogfuzz, the point in time when reaching the plateau varies from parser to parser: between 10 seconds (Pojo, Fig. \ref{fig:pojo}) and 450 seconds (json-simple, Fig. \ref{fig:jsonsimple}).
In contrary, for Rhino (Fig. \ref{fig:rhino}) both approaches cannot achieve a plateau within 10 minutes as they are able to continuously increase the line coverage over the time.

Table~\ref{table:evaluation} shows the accumulated coverage results for each subject and approach over all 30 repetitions.
For both approaches, Table~\ref{table:evaluation} shows the maximum and median line coverage, the standard deviation as well as the number of generated input files, along with the increase of the median line coverage of \evogfuzz compared to the baseline.
The improvement of the median line coverage ranges from 1.00\% (Pojo) to 47.93\% (Rhino).
Additionally, the standard deviation (SD) values for the baseline in Table \ref{table:evaluation} indicate the existence of plateaus because all repetitions for each JSON parser show a very low (and often 0\%) SD value.

\begin{table}[!t]
\begin{center}
\caption{Coverage results for each subject and approach over 30 repetitions.}
\label{table:evaluation}
\vspace{-1em}
\resizebox{\textwidth}{!}{
\begin{tabular}{| c | c | c | c | c | c | c | c | c | c | c | c | c |}
  \hline
  \multirow{2}{*}{Subject}&\multirow{2}{*}{LOC} &\multicolumn{4}{c|}{\evogfuzz} &\multicolumn{4}{c|}{Baseline} & Median & \multirow{2}{*}{p-value} \\
  & & max & median & SD & \#files & max & median & SD& \#files  & increase & \\
 \hline
 \hline
 ARGO & 8,265 & 49.78\% & 48.48\% & 0.60\% &11,900& 43.19\% & 43.19\% & 0\% & 13,900& 12.25\%& \textcolor{black}{0.000062}\\
Genson & 18,780 & 19.65\% & 19.09\% & 0.19\% & 8,100&  16.17\% & 16.17\% & 0\%  &9,000& 18.06\% & \textcolor{black}{0.000063}\\
Gson &25,172 & 26.92\% & 26.67\%& 0.15\% & 9,800&  24.08\% & 24.08\% & 0\%  & 11,200&10.76\%& \textcolor{black}{0.000064}\\
JSONJava&3,742 & 21.09\% & 18.47\% & 0.59\% & 12,700& 16.72\% & 16.72\% & 0\%  & 15,000&10.47\% & \textcolor{black}{0.000064}\\
JsonToJava & 5,131 & 18.58\% & 17.90\% & 0.39\%& 11,400& 17.58\% & 17.45\% & 0.09\%  &13,400& 2.58\% & \textcolor{black}{0.020699}\\
minimalJSON  & 6,350 & 51.06\% & 50.83\% & 0.26\% &14,000&  46.38\% &  46.38\% & 0\%  & 16,600&9.59\%& \textcolor{black}{0.000055}\\
Pojo &18,492 &  32.33\% & 32.17\%& 0.07\% & 10.600& 31,88\% & 31.88\% & 0.02\%  & 12,100& 1.00\% & \textcolor{black}{0.000061} \\
 json-simple& 2,432 & 34.44\% &  33.80\%& 0.33\%& 14,200&  28.54\% & 28.54\% & 0\%  & 16,700&18.43\%& \textcolor{black}{0.000059}\\
  \hline
   Rhino& 100,234 & 15.42\% & 13.95\% & 0.43\% &3,200  & 10.20\% & 9.43\% & 0.28\%  &3,800&47.93\%& \textcolor{black}{0.000183} \\
\hline
 cssValidator& 120,838 &7.53\% & 7.06\% & 0.21\% & 1,000 & 6.62\% & 6.51\% & 0.06\% & 2,500 & 8.45\%& \textcolor{black}{0.000183} \\
\hline
\end{tabular}
}
\end{center}
\vspace{-2em}
\end{table}

To support the graphical evaluation, we do a statistical analysis to increase the confidence in our conclusions.
As we consider independent samples and cannot make any assumption about the distribution of the results, we perform a non-parametric Mann-Whitney~U test~\cite{mann_test_1947,arcuri_hitchhikers_2014} to check whether the achieved median line coverage of both approaches differ significantly for each subject.
This statistical analysis confirms that \evogfuzz produces a significantly higher line coverage than the baseline for all subjects (cf. last column of Table~\ref{table:evaluation}).

The \#files columns in Table~\ref{table:evaluation} denote the average number of input files generated by one approach when testing one subject for 10 minutes. For all subjects, the baseline is able to generate on average more files than \evogfuzz. These differences indicate the costs of the evolutionary algorithm in \evogfuzz being eventually irrelevant due to the improved line coverage achieved by \evogfuzz.

\begin{figure}[t]
    \centering
    \begin{minipage}[b]{0.45\textwidth}
        \centering
        \includegraphics[width=48mm]{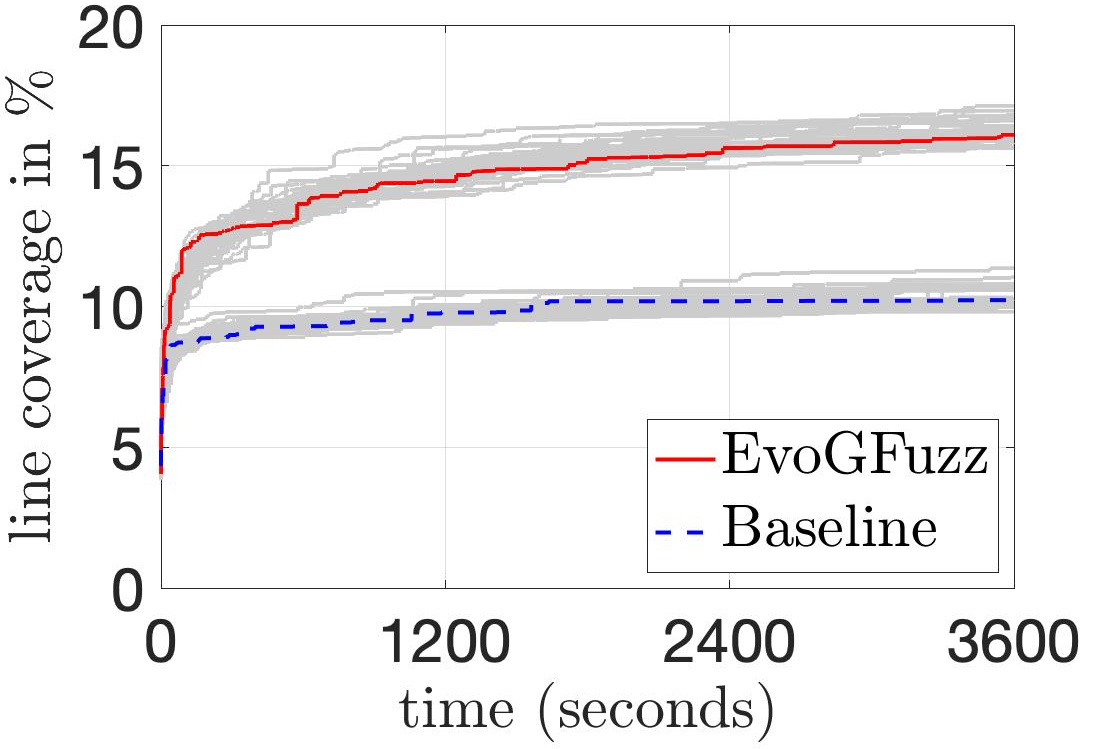} 
        \vspace{-1em}
        \caption{Line coverage of Rhino.}
        \label{fig:rhino_3600}
    \end{minipage}\hfill
    \begin{minipage}[b]{0.45\textwidth}
        \centering
        \includegraphics[width=40mm]{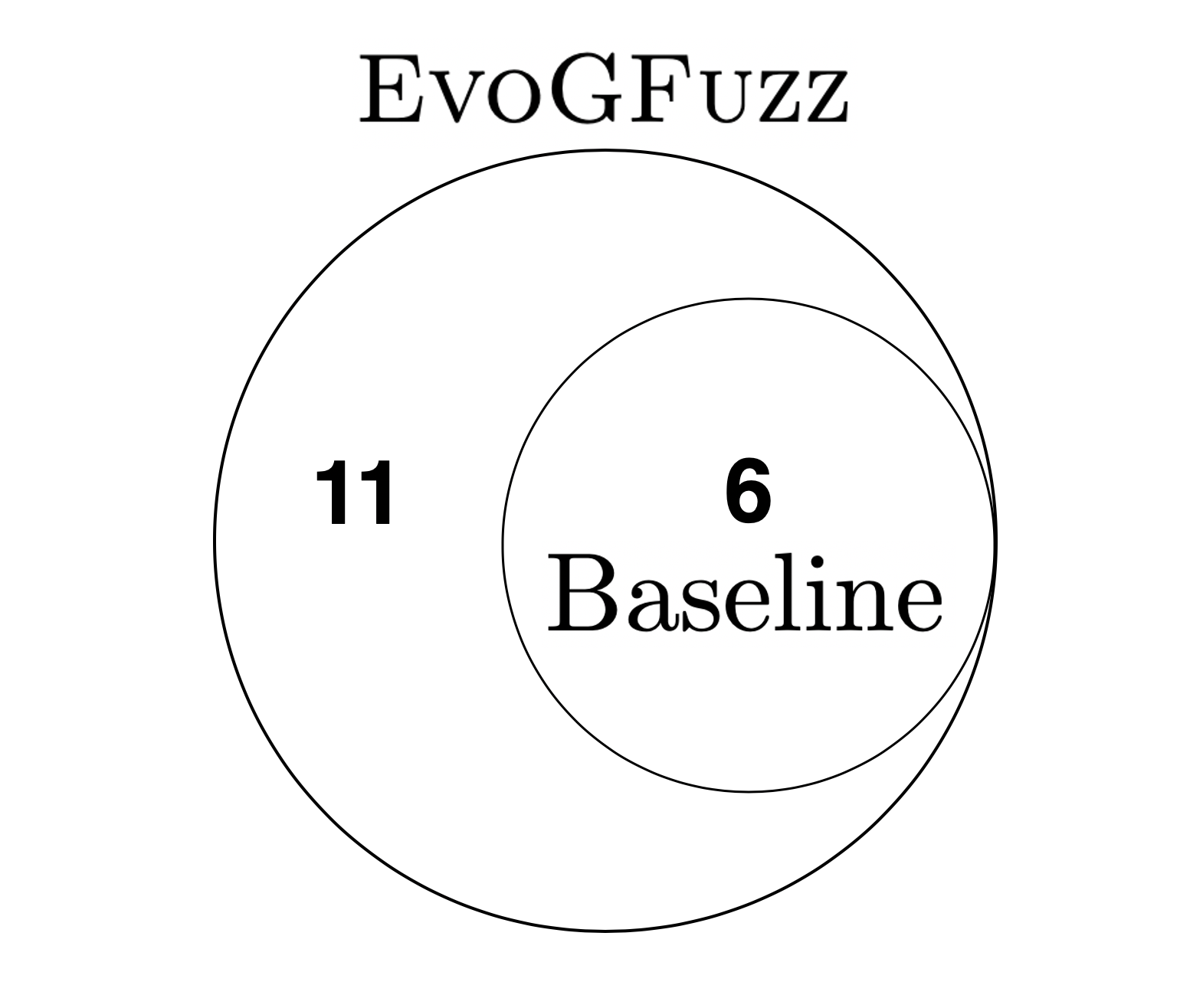}
        \vspace{-1em}
        \caption{Unique exceptions triggered by \evogfuzz (11) and the baseline (6).}
        \label{fig:overlap}
    \end{minipage}
	\vspace{-1em}
\end{figure}

Since both approaches managed to continuously increase the line coverage for the Rhino parser (Figure~\ref{fig:rhino}), we conducted an additional experiment with the time frame set to one hour and again repeated the experiment 30 times.
The results can be seen in Figure \ref{fig:rhino_3600}. The chart shows that both approaches managed to further improve their (median) line coverage. \evogfuzz was able to improve its previously achieved median line coverage of 13.95\% to 16.10\% with 18,500 generated input files, while the baseline improved from 9.43\% to 10.23\% with 22,700 generated input files, separating botch approaches even further.

\begin{tcolorbox}
Based on our evaluation, we conclude that \evogfuzz is able to achieve a \textbf{significantly higher} line coverage than the baseline.
\end{tcolorbox}

\begin{table}[!b]
\vspace{-1em}
\begin{center}
	\caption{Exception types that have been triggered by both approaches.}
	\label{table:exceptions}
	\vspace{-1em}
	\resizebox{.9\textwidth}{!}{
		\begin{tabular}{| c | c | c | @{\hspace{.5em}} r @{\hspace{.5em}} | @{\hspace{.5em}} r @{\hspace{.5em}} |}
			\hline
			Input & \multirow{2}{*}{Subject} & Exception & \multirow{2}{*}{\textsc{EvoGFuzz}}  & \multirow{2}{*}{Baseline}   \\
			language&  & types &  &  \\
			\hline
			\hline
			\multirow{9}{*}{JSON} & ARGO & argo.saj.InvalidSyntaxException &  \textcolor{blue}{\textbf{30}} / 30 &  \textcolor{red}{\textbf{0}} / 30\\
			&Genson &java.lang.NullPointerException & 30 / 30 & 30 / 30 \\
			&jsonToJava & org.json.JSONException & 30 / 30 & 30 / 30 \\
			&jsonToJava & java.lang.IllegalArgumentException & 30 / 30 & 30 / 30 \\
			&jsonToJava  & java.lang.ArrayIndexOutOfBoundsException & 30 / 30 & 30 / 30 \\
			&jsonToJava  & java.lang.NumberFormatException &  \textcolor{blue}{\textbf{6}} / 30 &  \textcolor{red}{\textbf{0}} / 30\\
			&Pojo & java.lang.StringIndexOutOfBoundsException& 30 / 30 & 30 / 30 \\
			& Pojo  & java.lang.IllegalArgumentException & 30 / 30 & 30 / 30 \\
			& Pojo  & java.lang.NumberFormatException &  \textcolor{blue}{\textbf{22}} / 30 &  \textcolor{red}{\textbf{0}} / 30\\
			\hline
			\multirow{2}{*}{JavaScript} & Rhino & java.lang.IllegalStateException &  \textcolor{blue}{\textbf{26}} / 30 &  \textcolor{red}{\textbf{0}} / 30\\
			& Rhino & java.util.concurrent.TimeoutException &  \textcolor{blue}{\textbf{15}} / 30 & \textcolor{red}{\textbf{0}} / 30\\
			\hline
			CSS3 & \multicolumn{4}{|c|}{No exceptions triggered}\\
			\hline
			\multicolumn{3}{|r|}{\textbf{Total exception types}}  &11 & 6   \\
			\hline
		\end{tabular}%
	}
\end{center}
\end{table}

\myparagraph{RQ2 - Exception Types}
To answer RQ2, we compare the number of times a unique exception type has been triggered.
Table~\ref{table:exceptions} shows the thrown exception types per subject and input language.
If neither approach was able to trigger an exception, the subject is not included in the table.
For the Gson, JsonJava, simple-json, minimal-json, and cssValidator parsers no defects and exceptions have been found by both approaches.
The ratios in the 4th and 5th column relate to the number of experiment repetitions in which \evogfuzz and the baseline were able to trigger the corresponding exception type.

Table \ref{table:exceptions} and Figure~\ref{fig:overlap} show that during each experiment repetition, \evogfuzz has been able to detect the same exception types than the baseline.
Furthermore, \evogfuzz was able to find five additional exception types that have not been triggered by the baseline.
However, apart from the exception type \emph{argo.saj.InvalidSyntaxException}, found in the ARGO parser, the other four exception types have not been identified by \evogfuzz in all repetitions.

\begin{tcolorbox}
Overall, \textbf{11 different exception types} in five subjects have been found in our evaluation, incl. just two custom types (\textit{org.json.JSONException} and \textit{argo.saj.InvalidSyntaxException}).
Out of these 11 exception types, \textbf{five} have \textbf{not} been triggered by the baseline.
Figure \ref{fig:overlap} shows that \textbf{all six} exception types triggered by the baseline were also found by \evogfuzz.
\end{tcolorbox}

\subsection{Threats to Validity}

\myparagraph{Internal Validity}
The main threats to internal validity of fuzzing evaluations are caused by the random nature of fuzzing \cite{arcuri_hitchhikers_2014,Klees2018EvaluateFuzzing}.
It requires a meticulous statistical assessment to make sure that observed behaviors are not randomly occurring.
Therefore, we repeated all experiments 30 times and reported the descriptive statistics of our results.
To match the evaluation of Pavese\,et\,al.~\cite{pavese_inputs_2018}, we used the same set of subjects and seed inputs.
Furthermore, we automated the data collection and statistical evaluation.
Finally, we did not tune the parameters of the baseline and \evogfuzz to reduce the threat of overfitting to the given grammars and subjects. Only for the fitness function of \evogfuzz, we determined appropriate $\lambda$ values for the three input grammars by experiments.

\myparagraph{External Validity}
The main threat to external validity is the generalizability of the experimental results that are based on a limited number of input grammars and systems under test.
However, similar to Pavese et al.~\cite{pavese_inputs_2018}, practically relevant input grammars with different complexities (small-sized grammars like JSON, and rather complex grammars like JavaScript and CSS) and widely used subjects (e.g., ARGO and Rhino) have been selected. As a result, we are confident that our approach will also work on other grammars and subjects.

\section{Conclusion and Future Work}\label{sec:conclusion} 
This paper presented \evogfuzz, \textit{evolutionary grammar-based fuzzing} that combines the technique by Pavese et al.~\cite{pavese_inputs_2018} with evolutionary optimization to direct the generation of complex and interesting inputs by a fitness function. \evogfuzz is able to generate structurally complex input files that trigger exceptions.
The introduced mutation of grammars maintains genetic diversity and allows \evogfuzz to discover features that have previously not been explored. Our experimental evaluation shows improved coverage compared to the original approach~\cite{pavese_inputs_2018}. Additionally, \evogfuzz is able to trigger more exception types undetected by the original approach. 
As future work, we want to investigate cases of having no precise grammar of the input space (cf.~\cite{Le2019Saffron}) and using semantic knowledge of the input language to tune mutation~operators.
Finally, we want to compare \evogfuzz with other state-of-the-art fuzzing techniques.

\bibliographystyle{splncs04}
\bibliography{evo-grammar-fuzzing,bib_ba}

\begin{thebibliography}{10}
\providecommand{\url}[1]{\texttt{#1}}
\providecommand{\urlprefix}{URL }
\providecommand{\doi}[1]{https://doi.org/#1}

\bibitem{Anand2013SurveyTesting}
Anand, S., Burke, E.K., Chen, T.Y., Clark, J., Cohen, M.B., Grieskamp, W.,
  Harman, M., Harrold, M.J., McMinn, P.: {An orchestrated survey of
  methodologies for automated software test case generation}. JSS
  \textbf{86}(8),  1978--2001 (2013)

\bibitem{arcuri_hitchhikers_2014}
Arcuri, A., Briand, L.: A {Hitchhiker}'s guide to statistical tests for
  assessing randomized algorithms in software engineering. Software Testing,
  Verification and Reliability  \textbf{24}(3),  219--250 (2014)

\bibitem{Atlidakis2020}
Atlidakis, V., Geambasu, R., Godefroid, P., Polishchuk, M., Ray, B.: {Pythia:
  Grammar-Based Fuzzing of REST APIs with Coverage-guided Feedback and
  Learning-based Mutations} pp. 1--12 (2020),
  \url{http://arxiv.org/abs/2005.11498}

\bibitem{Bohme2017AFLGo}
B\"{o}hme, M., Pham, V.T., Nguyen, M.D., Roychoudhury, A.: Directed greybox
  fuzzing. In: Proceedings of the ACM SIGSAC Conference on Computer and
  Communications Security. pp. 2329--2344. CCS '17, ACM (2017)

\bibitem{Bohme2016AFLFast}
B\"{o}hme, M., Pham, V.T., Roychoudhury, A.: Coverage-based greybox fuzzing as
  markov chain. In: Proceedings of the ACM SIGSAC Conference on Computer and
  Communications Security. pp. 1032--1043. CCS '16, ACM (2016)

\bibitem{Cummins2018}
Cummins, C., Petoumenos, P., Murray, A., Leather, H.: {Compiler fuzzing through
  deep learning}. In: Proceedings of the 27th ACM SIGSOFT International
  Symposium on Software Testing and Analysis. pp. 95--105. ISSTA 2018, ACM
  (2018)

\bibitem{du_elitism_2018}
Du, H., Wang, Z., Zhan, W., Guo, J.: Elitism and {Distance} {Strategy} for
  {Selection} of {Evolutionary} {Algorithms}. IEEE Access  \textbf{6},
  44531--44541 (2018)

\bibitem{Godefroid2020Fuzzing}
Godefroid, P.: Fuzzing: Hack, art, and science. Commun. ACM  \textbf{63}(2),
  70--76 (2020)

\bibitem{Godefroid2008GrammarWhiteBoxFuzzing}
Godefroid, P., Kiezun, A., Levin, M.Y.: Grammar-based whitebox fuzzing. In:
  Proceedings of the 29th ACM SIGPLAN Conference on Programming Language Design
  and Implementation. pp. 206--215. PLDI '08, ACM (2008)

\bibitem{Godefroid2012Sage}
Godefroid, P., Levin, M.Y., Molnar, D.: Sage: Whitebox fuzzing for security
  testing. Commun. ACM  \textbf{55}(3),  40--44 (2012)

\bibitem{Godefroid2017LearnFuzz}
Godefroid, P., Peleg, H., Singh, R.: {Learn{\&}Fuzz: Machine learning for input
  fuzzing}. In: Proceedings of the 32nd International Conference on Automated
  Software Engineering. pp. 50--59. ASE 2017, IEEE (2017)

\bibitem{hallaraker_detecting_2005}
Hallaraker, O., Vigna, G.: Detecting malicious {JavaScript} code in {Mozilla}.
  In: Proceedings of the 10th IEEE International Conference on Engineering of
  Complex Computer Systems. pp. 85--94. ICECCS '05, IEEE (2005)

\bibitem{Hanford1970}
{Hanford}, K.V.: Automatic generation of test cases. IBM Systems Journal
  \textbf{9}(4),  242--257 (1970)

\bibitem{Harman2012SBSE}
Harman, M., McMinn, P., de~Souza, J.T., Yoo, S.: Search based software
  engineering: Techniques, taxonomy, tutorial. In: Empirical Software
  Engineering and Verification: Intl. Summer Schools, LASER 2008-2010, pp.
  1--59. Springer (2012)

\bibitem{Holler2012FuzzingCodeFragments}
Holler, C., Herzig, K., Zeller, A.: Fuzzing with code fragments. In: Presented
  as part of the 21st {USENIX} Security Symposium. pp. 445--458. {USENIX}
  (2012)

\bibitem{Hoschele2017AUTOGRAM2}
{H\"{o}schele}, M., {Zeller}, A.: Mining input grammars with autogram. In: 39th
  International Conference on Software Engineering Companion. pp. 31--34. IEEE
  (2017)

\bibitem{Klees2018EvaluateFuzzing}
Klees, G., Ruef, A., Cooper, B., Wei, S., Hicks, M.: Evaluating fuzz testing.
  In: Proceedings of the 2018 ACM SIGSAC Conference on Computer and
  Communications Security. pp. 2123--2138. CCS '18, ACM (2018)

\bibitem{Le2019Saffron}
Le, X.B.D., P\u{a}s\u{a}reanu, C., Padhye, R., Lo, D., Visser, W., Sen, K.:
  {Saffron: Adaptive Grammar-Based Fuzzing for Worst-Case Analysis}. SIGSOFT
  Softw. Eng. Notes  \textbf{44}(4), ~14 (2019)

\bibitem{Lemieux2018FairFuzz}
Lemieux, C., Sen, K.: {FairFuzz: A Targeted Mutation Strategy for Increasing
  Greybox Fuzz Testing Coverage}. In: Proceedings of the 33rd ACM/IEEE
  International Conference on Automated Software Engineering. pp. 475--485.
  ASE, ACM (2018)

\bibitem{Liu2017}
Liu, P., Zhang, X., Pistoia, M., Zheng, Y., Marques, M., Zeng, L.: {Automatic
  Text Input Generation for Mobile Testing}. In: Proceedings of the 39th
  International Conference on Software Engineering. pp. 643--653. ICSE '17,
  IEEE (2017)

\bibitem{mann_test_1947}
Mann, H.B., Whitney, D.R.: On a {Test} of {Whether} one of {Two} {Random}
  {Variables} is {Stochastically} {Larger} than the {Other}. The Annals of
  Mathematical Statistics  \textbf{18}(1),  50--60 (1947)

\bibitem{Miller1990}
Miller, B.P., Fredriksen, L., So, B.: An empirical study of the reliability of
  unix utilities. Commun. ACM  \textbf{33}(12),  32--44 (1990)

\bibitem{miller_genetic_1995}
Miller, B.L., Goldberg, D.E.: Genetic {Algorithms}, {Tournament} {Selection},
  and the {Effects} of {Noise}. Complex Systems  \textbf{9},  193--212 (1995)

\bibitem{miller_systematic_1963}
Miller, J.C., Maloney, C.J.: Systematic mistake analysis of digital computer
  programs. Communications of the ACM  \textbf{6}(2),  58--63 (1963)

\bibitem{diffuzz}
Nilizadeh, S., Noller, Y., P\u{a}s\u{a}reanu, C.S.: Diffuzz: Differential
  fuzzing for side-channel analysis. In: Proceedings of the 41st International
  Conference on Software Engineering. pp. 176--187. ICSE '19, IEEE (2019)

\bibitem{Orso2014SoftwareTesting}
Orso, A., Rothermel, G.: Software testing: A research travelogue (2000-2014).
  In: Future of Software Engineering. pp. 117--132. FOSE 2014, ACM (2014)

\bibitem{Pacheco2007Randoop}
Pacheco, C., Ernst, M.D.: {Randoop: Feedback-directed random testing for Java}.
  In: Proceedings of the 22nd Conference on Object-Oriented Programming Systems
  and Applications Companion. pp. 815--816. OOPSLA '07, ACM (2007)

\bibitem{pavese_inputs_2018}
Pavese, E., Soremekun, E., Havrikov, N., Grunske, L., Zeller, A.: Inputs from
  {Hell}: {Generating} {Uncommon} {Inputs} from {Common} {Samples}.
  arXiv:1812.07525 [cs]  (2018), \url{http://arxiv.org/abs/1812.07525}

\bibitem{pham_smart_2019}
Pham, V.T., B\"{o}hme, M., Santosa, A.E., C\u{a}ciulescu, A.R., Roychoudhury,
  A.: Smart greybox fuzzing. IEEE Trans. on Software Engineering pp. 1--17
  (2019)

\bibitem{richardson_csi_2008}
Richardson, R.: {CSI} computer crime and security survey. Computer security
  institute  \textbf{1},  1--30 (2008)

\bibitem{Song2008Bitblaze}
Song, D., Brumley, D., Yin, H., Caballero, J., et~al.: Bitblaze: A new approach
  to computer security via binary analysis. In: Information Systems Security.
  pp. 1--25. Springer (2008)

\bibitem{veggalam_ifuzzer_2016}
Veggalam, S., Rawat, S., Haller, I., Bos, H.: {IFuzzer}: {An} {Evolutionary}
  {Interpreter} {Fuzzer} {Using} {Genetic} {Programming}. In: Computer
  {Security} – {ESORICS} 2016. pp. 581--601. Lecture {Notes} in {Computer}
  {Science}, Springer (2016)

\bibitem{Wang2019Superion}
Wang, J., Chen, B., Wei, L., Liu, Y.: {Superion: Grammar-Aware Greybox
  Fuzzing}. In: Proceedings of the 41st International Conference on Software
  Engineering. pp. 724--735. ICSE '19, IEEE (2019)

\bibitem{noauthor_afl_2018}
Website: {American} {Fuzzing} {Lop} ({AFL}) (2018),
  \url{http://lcamtuf.coredump.cx/afl/}

\bibitem{noauthor_libfuzzer_2018}
Website: {libFuzzer: A library for coverage-guided fuzz testing} (2018),
  \url{https://llvm.org/docs/LibFuzzer.html}

\bibitem{wright_evolution_1929}
Wright, S.: The {Evolution} of {Dominance}. The American Naturalist
  \textbf{63}(689),  556--561 (1929)

\bibitem{Yang2011CSmith}
Yang, X., Chen, Y., Eide, E., Regehr, J.: Finding and understanding bugs in c
  compilers. SIGPLAN Not.  \textbf{46}(6),  283--294 (2011)

\bibitem{fuzzingbook2019}
Zeller, A., Gopinath, R., B{\"o}hme, M., Fraser, G., Holler, C.: The fuzzing
  book. In: The Fuzzing Book. Saarland University (2019),
  \url{https://www.fuzzingbook.org/}

\end{thebibliography}

\end{document}